\documentclass[12pt]{article}
\usepackage[utf8]{inputenc}
\usepackage{authblk}
\usepackage{setspace}
\usepackage[margin=1.25in]{geometry}
\usepackage{graphicx}
\usepackage{subcaption}
\usepackage{amsmath}
\usepackage{lineno}
\usepackage{lipsum}
\usepackage[style=nejm, 
citestyle=numeric-comp,
sorting=none]{biblatex}
\begin{document}
\title{Identified hadron production at mid-rapidity in Au+Au collisions at $\sqrt{s_{NN}} = 54.4$ GeV at STAR}
\author[1,*]{Krishan Gopal (for the STAR Collaboration)}
\affil[1]{Department of Physics, Indian Institute of Science Education and Research Tirupati}
\affil[*]{krishangopal@students.iisertirupati.ac.in}

\onehalfspacing
\maketitle

\date{}


\begin{abstract}

Quantum Chromodynamics (QCD) predicts that at sufficiently high temperature ($T$) and/or baryon chemical potential ($\mu_B$), the state of matter is in the form of quarks and gluons, which are no longer confined within hadrons. This deconfined state of matter is known as the Quark-Gluon Plasma (QGP). The goal of relativistic heavy-ion collision experiments is to create such a hot and dense state of matter and study its properties. Measurements of identified particle spectra in Au+Au collisions provide information on the bulk properties, such as integrated yield ($dN/dy$), average transverse momenta ($\langle p_T \rangle$), particle ratios, and freeze-out parameters of the medium produced. The systematic study of bulk properties sheds light on the particle production mechanism in these collisions. Also, the centrality dependence of the freeze-out parameters provides an opportunity to explore the QCD phase diagram.

In this talk, we will present the transverse momentum spectra of identified hadrons ($\pi^{\pm}$, $K^{\pm}$, $p$, and $\bar{p}$) at mid-rapidity ($|y| < 0.1$) in Au+Au collisions at $\sqrt{s_{\mathrm{NN}}} = 54.4$ GeV. The centrality dependence of $dN/dy$, particle ratios, and kinetic freeze-out parameters will also be presented, and their physics implications will be discussed. In addition, we will compare our results with previously published results at other collision energies.
\end{abstract}
\section{Introduction}
Quantum Chromodynamics (QCD) predicts the formation of the Quark-Gluon Plasma (QGP), a new state of matter, in heavy-ion collisions at high energy density or temperature \cite{RefA}. Studying transverse momentum spectra in heavy-ion collisions provides crucial information on QGP bulk properties, contributing to our understanding of the QCD phase diagram, particle production mechanisms, and freeze-out properties of the created medium. In this report, we present the transverse momentum spectra of identified hadrons in Au+Au collisions at $\sqrt{s_{NN}} = 54.4$ GeV using the Time Projection Chamber (TPC) and Time of Flight (TOF) detectors at STAR.

\section{Results and Discussions}
\begin{figure}[htbp]
\centering
\includegraphics[width=15.0cm,height=4.850cm]{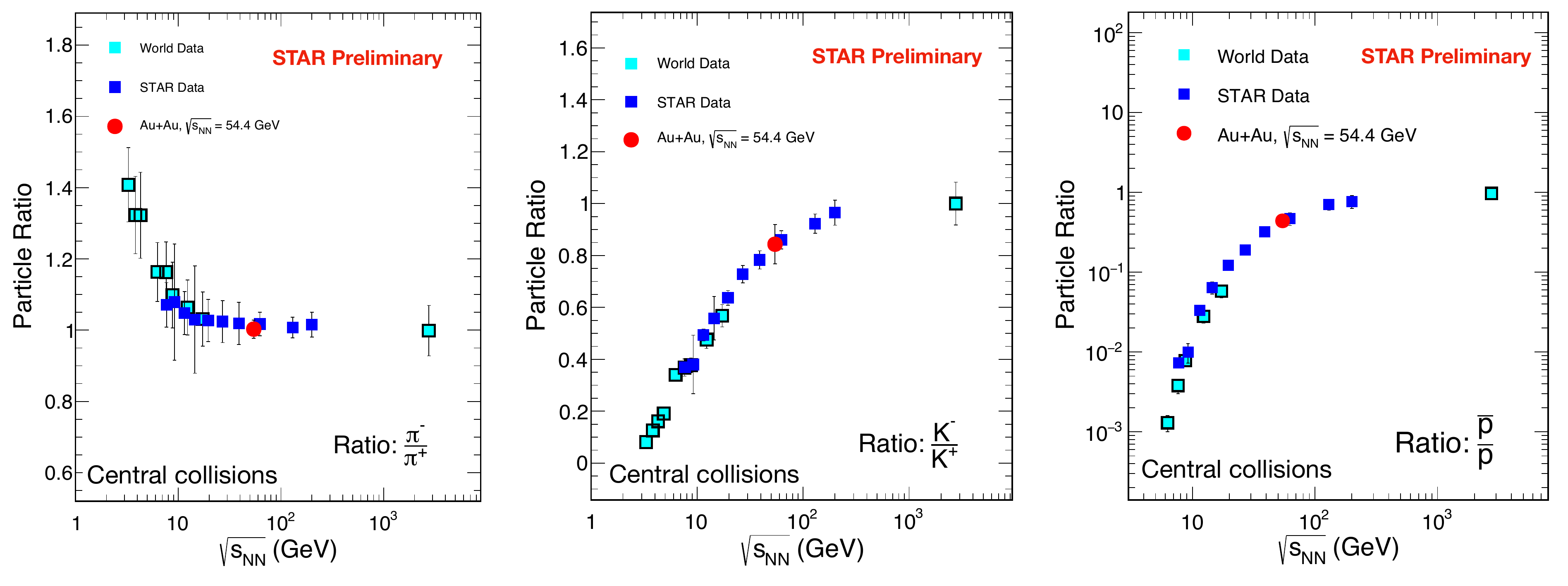}
\caption{$\pi^{-}/\pi^{+}$, $K^{-}/K^{+}$, and $\bar{p}/p$ ratios at mid-rapidity ($|y| < 0.1$) in 0--5\% Au+Au collisions at $\sqrt{s_{NN}}$ = 7.7--200 GeV. The uncertainties are statistical and systematic added in quadrature.}
\label{fig:figure1}
\end{figure}

Figure \ref{fig:figure1} shows particle ratios ($\pi^-/\pi^+$, $K^-/K^+$, and $\bar{p}/p$) in the most central (0-5\%) collisions as a function of collision energy. At lower beam energies, the $\pi^-/\pi^+$ ratios exceed unity due to the contributions from resonance decays like $\Delta$ baryons. The $K^-/K^+$ ratios show an increasing trend with increasing $\sqrt{s_{NN}}$ and approaches unity at higher beam energies, signifying the associated production of $K^+$ at lower energies. The $\bar{p}/p$ ratios increase with increasing $\sqrt{s_{NN}}$ but approach unity at the highest RHIC energy, indicating stronger baryon stopping at lower energies. The 54.4 GeV results follow the trend shown from previous measurements \cite{lokesh} of AGS, SPS, RHIC, and LHC. 

\begin{figure}[htbp]
\centering
\includegraphics[width=7.0cm,height=5.850cm]{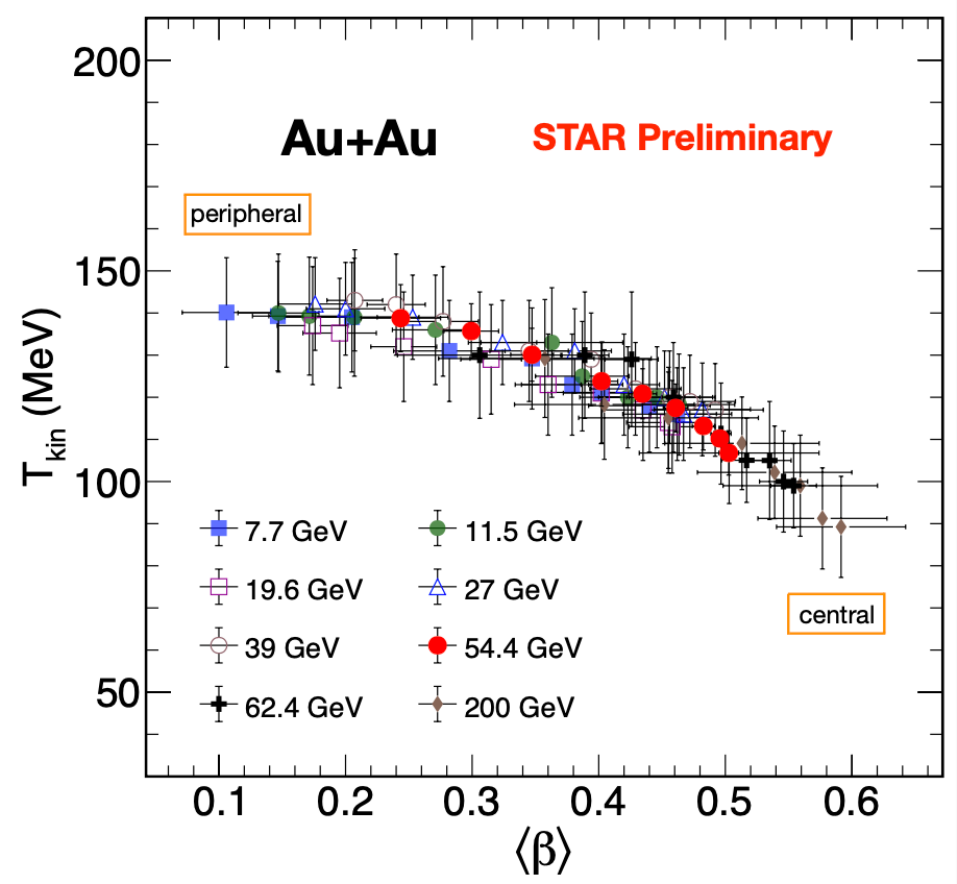}
\caption{Variation of $T_{kin}$ with $\langle \beta \rangle$ for various centralities in different collision energies.}
\label{fig:figure2}
\end{figure}
 
A simultaneous fit to the $p_T$ spectra of $\pi$, $K$, $p$, and their antiparticles was performed in different centrality intervals for Au+Au collisions at $\sqrt{s_{NN}} = 54.4$ GeV using the blast-wave model \cite{bw1,bw2} to study the kinetic freeze-out properties of the medium. Figure \ref{fig:figure2} shows that as we move from central to peripheral collisions, there is a decrease in transverse flow velocity ($\langle \beta \rangle$) and an increase in kinetic freeze-out temperature ($T_{kin}$), consistent with the expectation of a shorter lived fireball towards peripheral collisions \cite{heinz}.
\printbibliography

\end{document}